\documentclass[aps,prl,superscriptaddress,citeautoscript,twocolumn,reprint,longbibliography,floatfix]{revtex4-2}
\usepackage{graphicx,amssymb,amsmath,epsf,bm,cprotect,comment,physics,siunitx}
\epsfclipon

\renewcommand{\eqref}[1]{Eq.\,(\ref{#1})}

\newcommand{\figpref}[2]{Fig.\,\ref{#1}(#2)}

\usepackage[labelfont=bf,justification=raggedright,font=small]{caption}
\DeclareCaptionLabelSeparator{vline}{ $\bm{|}$ }
\usepackage{xr} 
\makeatletter
\newcommand*{\addFileDependency}[1]{
  \typeout{(#1)}
  \@addtofilelist{#1}
  \IfFileExists{#1}{}{\typeout{No file #1.}}
}
\makeatother

\newcommand*{\myexternaldocument}[1]{
    \externaldocument[S-]{#1}
    \addFileDependency{#1.tex}
    \addFileDependency{#1.aux}
}
\myexternaldocument{suppl}

\begin{document}

\title{Designing topological edge states in bacterial active matter}

\author{Yoshihito Uchida}
\email{uchida@noneq.phys.s.u-tokyo.ac.jp}
\affiliation{Department of Physics,\! The University of Tokyo,\! 7-3-1 Hongo,\! Bunkyo-ku,\! Tokyo 113-0033,\! Japan}%

\author{Daiki Nishiguchi}
\email{nishiguchi@phys.sci.isct.ac.jp}
\affiliation{Department of Physics,\! Institute of Science Tokyo,\! 2-12-1 Ookayama,\! Meguro-ku,\! Tokyo 152-8551,\! Japan}%
\affiliation{Department of Physics,\! The University of Tokyo,\! 7-3-1 Hongo,\! Bunkyo-ku,\! Tokyo 113-0033,\! Japan}%

\author{Kazumasa A. Takeuchi}
\email{kat@kaztake.org}
\affiliation{Department of Physics,\! The University of Tokyo,\! 7-3-1 Hongo,\! Bunkyo-ku,\! Tokyo 113-0033,\! Japan}%
\affiliation{Universal Biology Institute,\! The University of Tokyo,\! 7-3-1 Hongo,\! Bunkyo-ku,\! Tokyo 113-0033,\! Japan}%
\affiliation{Institute for Physics of Intelligence,\! The University of Tokyo,\! 7-3-1 Hongo,\! Bunkyo-ku,\! Tokyo 113-0033,\! Japan}%

\date{\today}

\begin{abstract}
Topology provides a unifying framework for understanding robust transport through protected edge states arising from nontrivial wavenumber topology. Extending these concepts to active matter, however, remains largely unexplored experimentally, with realizations limited to systems composed of chiral active particles. Here, we realize topological edge states in dense bacterial suspension, which represents a prototypical active matter system, using microfabricated geometrical structures with nontrivial wavenumber topology. Inspired by previous theoretical studies, we constructed a directional kagome network composed of ratchet-shaped channels that induce unidirectional bacterial flow. In this network, we found clear edge localization of bacterial density. A steady-state analysis based on the bacterial transport model and experimentally measured velocity field reveals how the characteristic collective flow generates edge localization. The model also uncovers the topological origin of the observed edge states. By tuning the geometry of the microfabricated networks, we identified directional channel design and network chirality as the key design features essential for the emergence of the edge state. Our results pave the way for establishing a control and design principle of topological transport in such active matter systems.
\end{abstract}

\maketitle

Topology offers a powerful framework for understanding robust transport phenomena in condensed matter systems, where nontrivial wavenumber topology gives rise to robust edge states. Representative examples include the quantum Hall effect \cite{PhysRevLett.45.494, PhysRevLett.49.405} and topological insulators \cite{RevModPhys.82.3045}, which link bulk topology to protected edge states through bulk-edge correspondence \cite{PhysRevLett.71.3697}. In recent years, these concepts have been applied beyond solid-state systems to a wide variety of physical systems, including living systems \cite{doi:10.1073/pnas.1721096115, PhysRevX.11.031015}. Indeed, it is intriguing to consider the possibility that topology contributes to the remarkable robustness often observed in the collective dynamics and functions of living systems. 

Building on this idea, active matter, consisting of self-propelled particles that convert energy into motion by themselves, offers a distinct platform for exploring topological phenomena. Theoretical studies have predicted that such systems can host topological edge states despite being intrinsically nonequilibrium and fluctuating \cite {10.1038/s42254-022-00445-3, sone2024hermitiannonhermitiantopologyactive}. One approach to realizing these states relies on chiral active particles, which generate edge flow, that is, strong unidirectional flow along the system boundary \cite{PhysRevLett.122.128001, 10.1038/s41467-020-19488-0}. Experimentally, edge flows have indeed  been observed in systems composed of active particles exhibiting self-spinning or chiral motion \cite{PhysRevE.101.022603, PhysRevLett.126.198001, 10.1038/s41567-019-0603-8, yamauchi2020chiralitydrivenedgeflownonhermitian, PhysRevX.14.041006}. An equally fundamental route is to realize topological phenomena through geometric design, where environmental structure rather than particle chirality induces directional transport \cite{10.1038/nphys4193, PhysRevLett.123.205502}. This strategy remains largely unexplored experimentally, yet it offers distinct advantages: geometric design enables systematic control by imposing periodic network structures, analogous to crystal lattices in solid-state systems, onto prototypical active matter. Such an approach provides a pathway to engineer new topological states in a controlled manner, demonstrating that even generic active matter systems can exhibit such robust behavior.

To experimentally explore this geometrical route, we engineer microfabricated networks with nontrivial wavenumber topology, drawing on theoretical predictions of edge states in nonequilibrium and stochastic systems \cite{10.1038/nphys4193, PhysRevLett.123.205502, PhysRevLett.125.258301, PhysRevE.104.025003,10.1038/s41598-021-04178-8, doi:10.1073/pnas.1721096115, PhysRevX.11.031015}, and use dense bacterial suspension as a versatile active matter platform. Dense bacterial suspension is ideal because its collective dynamics can be strongly influenced by microfabricated geometries. In unconfined environments, it exhibits turbulent-like flows known as active turbulence \cite{Aranson_2022, annurev:/content/journals/10.1146/annurev-conmatphys-082321-035957}, whereas microfabricated geometries can induce distinct transport behaviors and ordered states. For example, an asymmetric gear immersed in bacterial suspension exhibits unidirectional rotation  \cite{diLeonardo2010RatchetMotors, doi:10.1073/pnas.0913015107}. Confinement also plays a central role in organizing collective behavior: in isolated circular wells, dense bacterial suspension spontaneously forms a single stable vortex, while periodic lattices of such wells can couple these vortices into ordered states analogous to magnetic spin systems \cite{annurev:/content/journals/10.1146/annurev-conmatphys-031016-025522, PhysRevLett.110.268102, doi:10.1073/pnas.1405698111, 10.1038/nphys3607, C7SM00999B, doi:10.1073/pnas.2107461118, 10.1038/s41467-018-06842-6,10.1038/s42005-020-0337-z, doi:10.1073/pnas.2414446122}. This versatility makes bacterial suspension a broadly useful platform for testing how geometry can be harnessed to induce topological edge states.

Here, inspired by theoretical models \cite{PhysRevE.104.025003, PhysRevLett.123.205502}, we experimentally realize topological edge states in dense bacterial suspension using microfabricated kagome networks with directional channels. In these networks, we reveal that rectified bacterial flows generated by ratchet-shaped channels give rise to edge localization of bacterial density, reminiscent of topological transport. To elucidate how these characteristic bacterial flows lead to the emergence of edge localization, we determine the steady-state solution of the bacterial transport equation by solving its eigenvalue problem, using the experimentally obtained velocity field. We further establish the topological origin of the observed edge localization, through a theoretical framework that describes bacterial transport on directional kagome networks and characterizes their wavenumber topology. Finally, by tuning the network geometry, we identify directional channel design and network chirality as the key structural features essential for the emergence of edge states, thereby demonstrating how environmental geometry can be designed to achieve topological transport in prototypical active matter. These findings open new avenues for realizing and controlling topological phenomena in active matter systems through geometrical design rather than particle-level modification.

\vspace{5mm}

\section*{Edge localization in a directional kagome network}


\begin{figure*}[p!]
\centering
\includegraphics[width=\hsize]{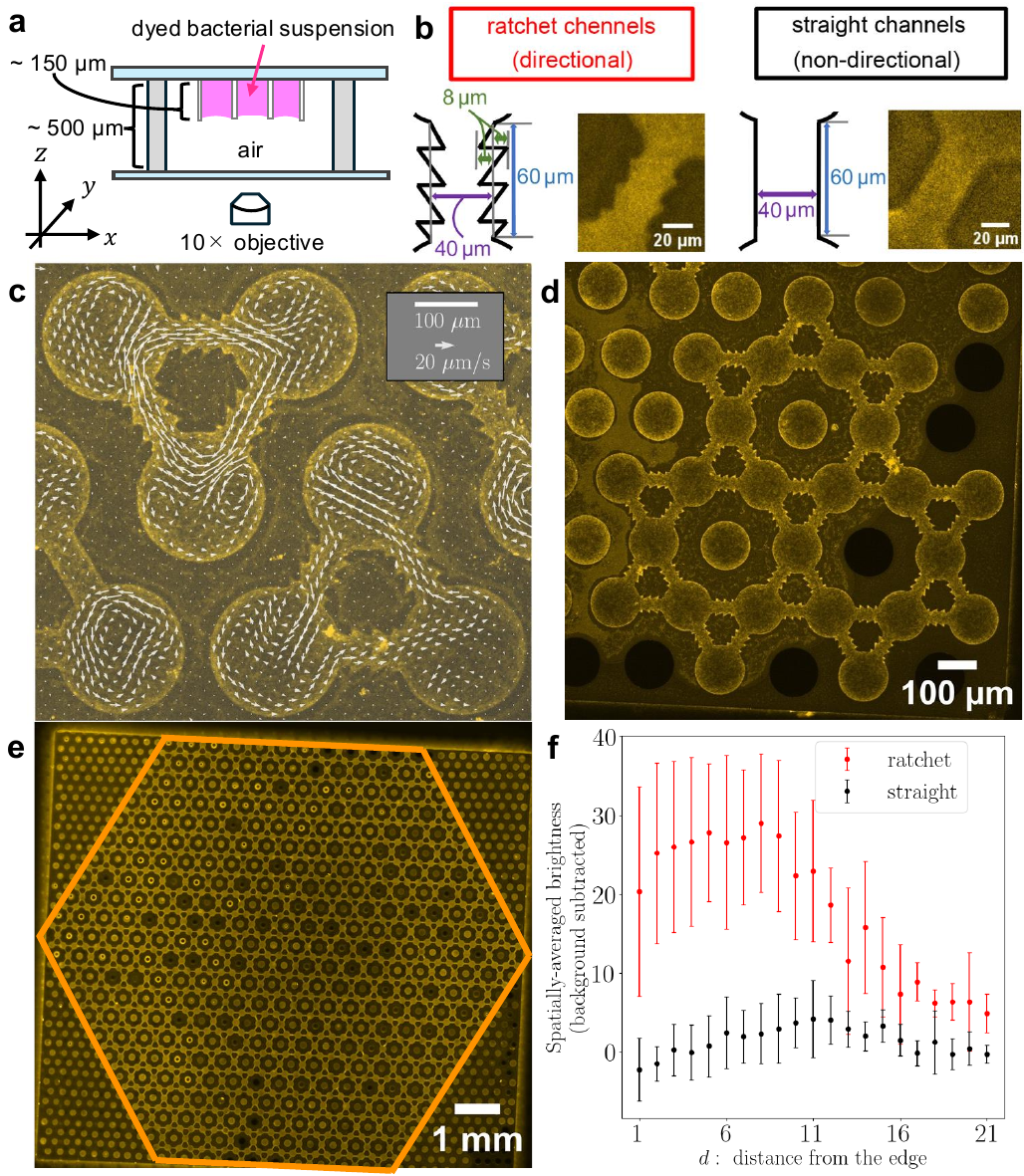}
\caption{
{\bf Edge localization in a directional kagome network.} {\bf a}, Experimental setup. Bacterial collective motion is observed near the liquid-air interface. {\bf b}, Schematics of directional ratchet-shaped and non-directional straight channels. Wide-field images of the channels filled with dyed bacterial suspension are also shown. {\bf c, d}, Confocal images of dyed bacterial suspension confined in triangular structures ({\bf c}) and a directional kagome network ({\bf d}). In the panel {\bf c}, time-averaged velocity field obtained by PIV analysis is overlaid. The length of the arrow in the box corresponds to a speed of $\SI{20}{\micro m/s}$. {\bf e}, Wide-field image of a large directional kagome network filled with dyed bacterial suspension, captured using epifluorescence microscopy with a tiling scan. The network boundary is indicated by orange lines. {\bf f}, Bacterial density profiles in the directional and non-directional kagome networks. Background-subtracted, spatially-averaged fluorescence intensity at wells along diagonal lines of the kagome network is plotted (see also Supplementary Fig.~1). The horizontal axis indicates the distance from the edge: $d=1$ corresponds to wells closest to the vertex of the orange hexagon in {\bf e}, while $d=21$ corresponds to wells in the central triangular motif of the network. The error bars represent the standard error over six data obtained from different diagonal lines.
}
\label{fig1}
\end{figure*}

To realize a system exhibiting topological transport phenomena in active matter systems, we conducted experiments using a suspension of \textit{Bacillus subtilis} (strain 1085), confined within microfabricated structures, as shown in \figpref{fig1}{a}. Owing to aerotactic behavior, bacteria concentrate near the liquid-air interface, where collective motion is predominantly observed. 

As a first step toward realizing geometry-driven topological transport, we sought to establish that bacterial collective motion can be directionally rectified through geometrical design, since such rectification serves as the fundamental mechanism for inducing directional flow and edge localization in extended networks. To this end, we fabricated structures consisting of circular wells (radius $r=\SI{65}{\micro m}$) connected by directional ratchet-shaped channels (length $l=\SI{60}{\micro m}$, mean width $w=\SI{40}{\micro m}$, tooth height $h=\SI{8}{\micro m}$), see \figpref{fig1}{b}. To test whether this design indeed induces unidirectional flow, we constructed triangular arrangements of wells linked by the ratchet-shaped channels and observed the collective motion of fluorescently-labeled bacteria, see \figpref{fig1}{c} and Supplementary Movie 1. In \figpref{fig1}{c}, the channel orientations were designed to favor clockwise flow on the left and counterclockwise flow on the right. Bacterial flow was imaged using an inverted microscope (IX83, Olympus) equipped with a spinning-disk confocal unit (Dragonfly 200, Andor), which is well suited for capturing fast collective behavior clearly. The time-averaged velocity field, shown in \figpref{fig1}{c}, reveals the emergence of unidirectional flow in the expected direction. These results demonstrate that the orientation of the directional channels effectively controls the direction of bacterial collective flow.

Next, inspired by the geometry of the theoretical models \cite{PhysRevE.104.025003, PhysRevLett.123.205502}, we constructed directional kagome networks, see e.g., \figpref{fig1}{d}. To clearly visualize the spatial distribution of bacterial density, we fabricated kagome networks consisting of 1023 circular wells and observed the steady-state density using epifluorescence microscopy with a tiling scan, see \figpref{fig1}{e}. To quantify the extent of bacterial density localization toward the edge, we calculated the spatially-averaged fluorescence intensity at each circular well. We then plotted the background-subtracted brightness along six diagonal lines of the kagome network, see red plots in \figpref{fig1}{f} and Supplementary Fig.~1. The clear observation of edge localization, manifested as a decay in bacterial density from the edge into the bulk, reveals a boundary-localized collective state, reminiscent of topological edge phenomena, in dense bacterial suspension.

We further tested whether the observed edge localization originates from the directional network geometry, by comparing bacterial density profiles in two types of kagome networks: one composed of directional ratchet-shaped channels (left panel of \figpref{fig1}{b}) and the other with non-directional straight channels (right panel of \figpref{fig1}{b}). As shown in \figpref{fig1}{f}, edge localization emerges only in the directional kagome network, whereas bacterial density remains uniformly distributed in the non-directional case, see also Supplementary Fig.~2. This comparison indicates that the rectified collective flow induced by the directional channels is essential for the emergence of edge localization.

\vspace{5mm}

\section*{Characteristic bacterial flow generates edge localization}

\begin{figure*}[p!]
\centering
\includegraphics[width=\hsize]{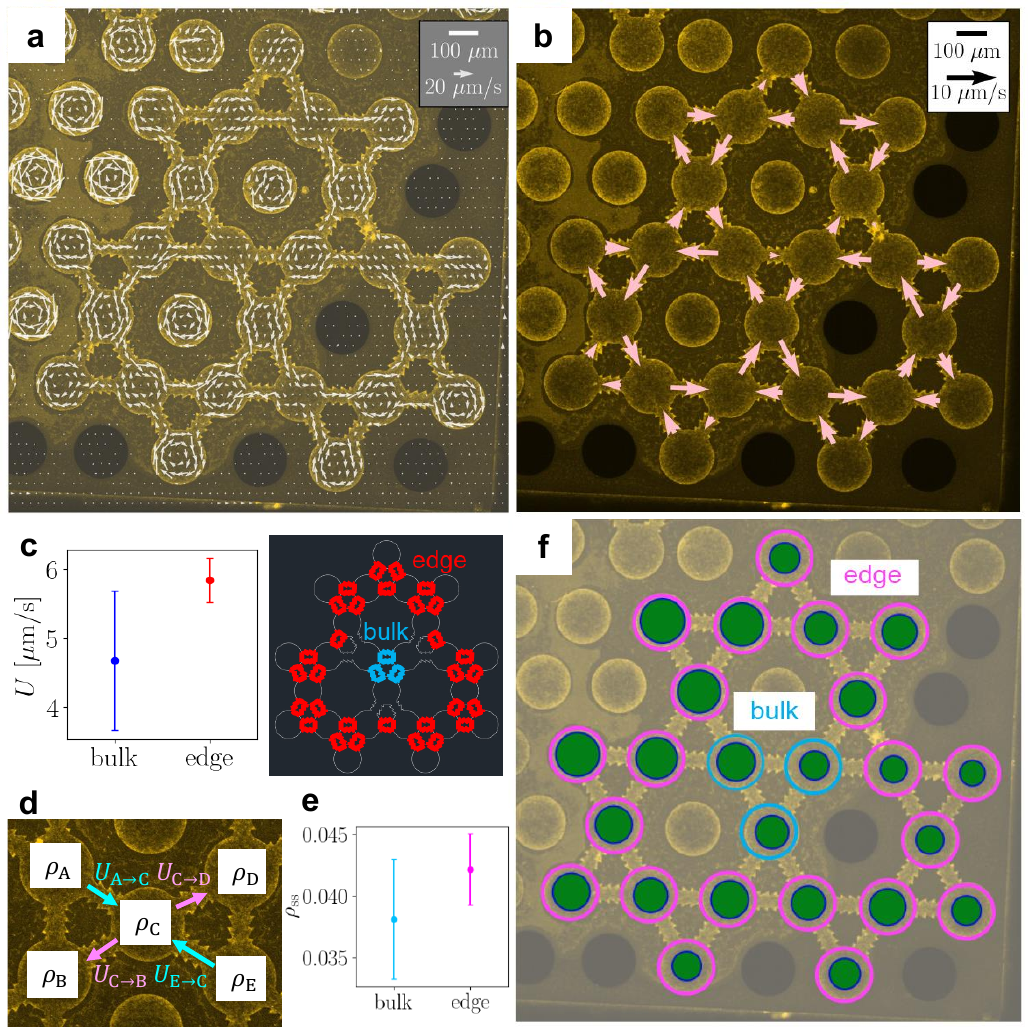}
\caption{{\bf Bacterial transport driven by characteristic collective flow.} {\bf a}, Time-averaged velocity field of bacterial collective flow in a small kagome network. Velocity vectors are overlaid on a confocal image of fluorescently labeled, dense bacterial suspension. {\bf b}, Map of channel flow $U$ calculated from the velocity field in each channel. The length of the arrow in the box corresponds to a speed of $\SI{10}{\micro m/s}$. {\bf c}, Comparison of channel flow $U$ in the bulk and along the edge. We focused on two types of channels in the small kagome network: those connecting wells within the triangular motif at the center of the network (blue, labeled \textit{bulk}) and those connecting wells along the edge (red, labeled \textit{edge}). The mean channel flow for each type is shown, with error bars indicating the standard error. {\bf d}, Schematic of the bacterial transport driven by the channel flow. Bacteria enter and leave a well through inlet and outlet channels, respectively. The time evolution of the density at the well labeled “C” in the figure is given by $\partial_t\rho_{\text{C}}=\rho_{\text{A}} U_{\text{A}\to\text{C}}+\rho_{\text{E}} U_{\text{E}\to\text{C}}-\rho_{\text{C}}U_{\text{C}\to\text{B}}-\rho_{\text{C}} U_{\text{C}\to\text{D}}$ [\eqref{eq1}]. {\bf e, f}, Steady-state bacterial density distribution, evaluated from the channel flow pattern (see Methods). In {\bf f}, the area of each green circle represents the bacterial density at that site. Two types of wells were analyzed: one type at the center of the network (blue, labeled \textit{bulk}) and the other along the edge (pink, labeled \textit{edge}). In {\bf e}, the mean steady-state bacterial density $\rho_{\text{ss}}$ for each type is shown, with error bars indicating the standard error.}
\label{fig2}
\end{figure*}

To elucidate how edge localization arises from the characteristic bacterial flow in directional kagome networks, we analyzed steady-state transport based on experimentally obtained velocity fields. So far, edge localization was identified from the steady-state bacterial density distribution in the large network composed of 1023 wells shown in \figpref{fig1}{e}. However, direct measurement of the instantaneous velocity fields in such a large system is technically infeasible due to the limited temporal resolution of a tiling scan. We therefore constructed a smaller directional kagome network consisting of 24 circular wells, which allowed us to simultaneously observe the collective flow throughout the entire network, see \figpref{fig1}{d}, \figpref{fig2}{a}, and Supplementary Movie 2.

We analyzed the channel flow $U$, defined as the velocity field averaged over both time and space within each channel, to quantify bacterial collective flow in ratchet-shaped channels. First, the time-averaged bacterial flow was obtained using the particle image velocimetry (PIV) analysis with PIVlab \cite{Thielicke-2021}. We then calculated the spatially-averaged velocity within each channel region and projected it onto the channel axis to obtain the channel flow, see \figpref{fig2}{b} and Supplementary Fig.~3. Notably, channels in the edge region tend to have stronger flow than those in the bulk, see \figpref{fig2}{c}. 

To examine whether the characteristic bacterial flow observed in the directional kagome network can account for the emergence of edge localization, we evaluated the steady-state bacterial density profile driven by the experimentally obtained channel flows, assuming the following equation for bacterial transport:
\begin{equation}
   \frac{\partial\rho_{I}}{\partial t}=\sum_{J\in\text{in}}\rho_{J}U_{J\to I}-\sum_{J\in\text{out}}\rho_{I}U_{I\to J}. \label{eq1} 
\end{equation}
Here, $\rho_I$ represents the bacterial density at site $I$, and $U_{I\to J}$ denotes the channel flow from site $I$ (the departure site) to site $J$ (the arrival site), see \figpref{fig2}{d}. The first and second terms in \eqref{eq1} represent the total inflow to and outflow from site $I$, respectively. If all channel flows were uniform, $U_{I\to J}=U_0$ (a constant), the analytical steady-state solution would yield a uniform density distribution across the network. In contrast, when we use the experimentally obtained, spatially varying $U_{I\to J}$ shown in \figpref{fig2}{b}, the steady-state solution shows that bacterial density tends to be higher along the edge than in the bulk, see \figpref{fig2}{e} and \figpref{fig2}{f}. These results indicate that the characteristic bacterial flow in the directional kagome network gives rise to the observed edge localization.

\vspace{5mm}

\section*{Theoretical characterization of the observed edge state}

\begin{figure*}[p!]
\centering
\includegraphics[width=\hsize]{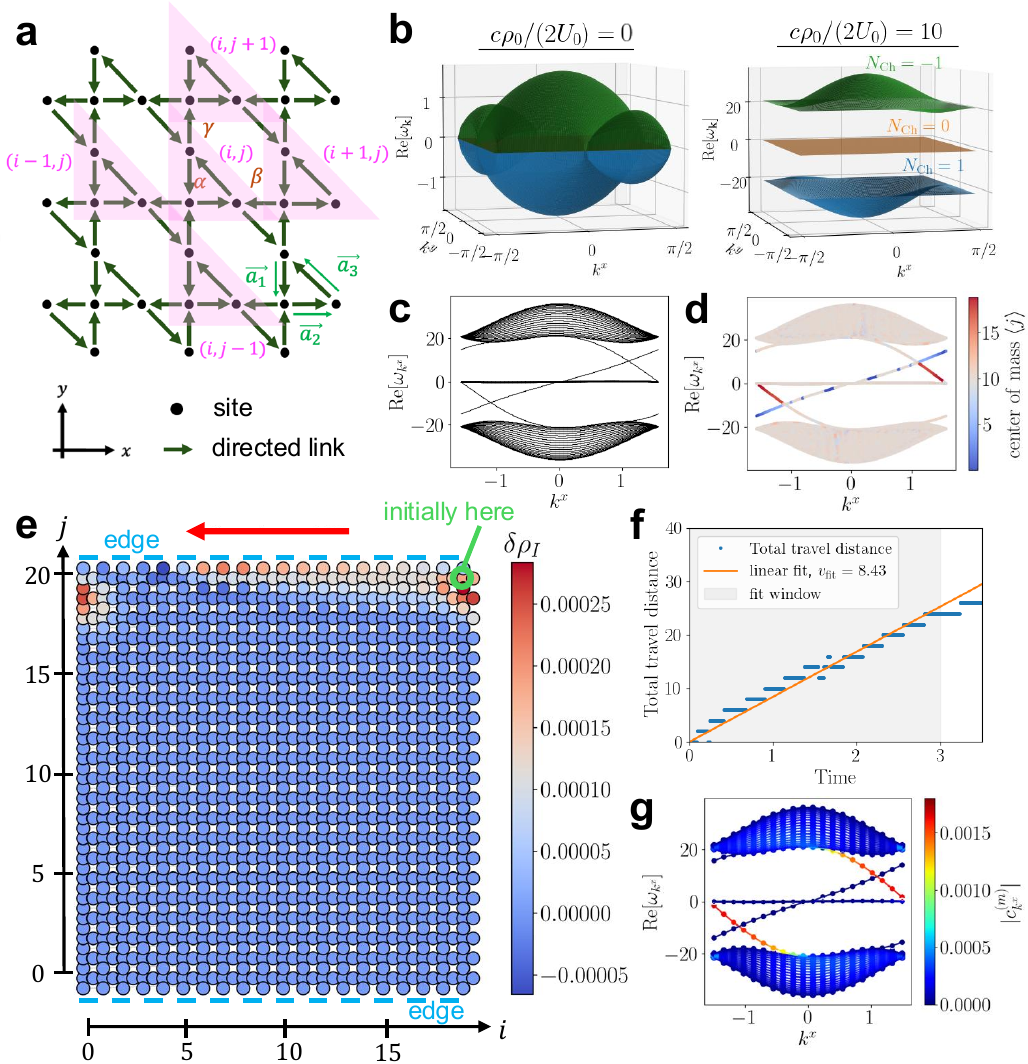}
\caption{{\bf Theoretical characterization of bacterial transport on the directional kagome network.} {\bf a}, Schematic of the directional kagome network. Each unit triangular motif, composed of three sites $\nu=\alpha,\beta,\gamma$, is labeled by $(i,j)$. {\bf b}, Bulk band structure under the fully periodic boundary condition. For $c>0$, the bands are separated by gaps, and the Chern numbers of the top, middle, and bottom bands are $-1$, $0$, and $1$, respectively. {\bf c}, Dispersion relation in the half-periodic geometry for $c\rho_0/(2U_0)=10$. {\bf d}, $y$-coordinate of the center of mass $\langle j\rangle$ for each eigenmode in the half-periodic geometry for $c\rho_0/(2U_0)=10$. Gap-crossing modes have their centers of mass concentrated near the bottom ($y\approx 0$) and top ($y\approx 20$) rows of the directional kagome network. {\bf e}, Snapshot at time $t=3.0$ from the direct simulation of the nonlinear model in the half-periodic geometry for $c\rho_0/(2U_0)=10$. The color shows the density deviation $\delta\rho_I=\rho_I-\rho_{0,I}$. At $t=0$, the density deviation $\delta\rho_I$ is set to 0.01 at the top-right corner, marked in green, and $\delta\rho_I=0$ at all other sites. The red arrow indicates the direction of the observed edge transport. See also Supplementary Movie 3. {\bf f}, Total travel distance $D(t)$ of the density peak along the top triangular motifs. The travel distance is defined as the distance from the triangular motif where the initial density deviation was applied (marked in green in {\bf e}). The orange line shows a linear fit, $D(t)=v_{\text{fit}} t$. The fit is performed over the time interval $t<3$ (shaded in gray), during which the density peak propagates at an approximately constant speed. From this fit, we obtain $v_{\text{fit}}=8.4\pm 1.2$. {\bf g}, Expansion coefficients $|c^{(m)}_{k^x}|$ at $t=0$ obtained by projecting density deviation in the simulation onto the eigenmodes of the half-periodic Hamiltonian for $c\rho_0/(2U_0)=10$.}
\label{fig3}
\end{figure*}

To further characterize the experimentally observed edge localization (\figpref{fig1}{e,f}, \figpref{fig2}{e,f}) and the stronger flow along the edge (\figpref{fig2}{c}) in the directional kagome network, we consider a simplified bacterial transport model. The time evolution of the bacterial density $\rho_{I}(t)$ at site $I$ is governed by \eqref{eq1}. To close the equation, we introduce a relationship between the channel flow $U_{I \to J}$ and the local bacterial densities $\rho_I(t)$ and $\rho_J(t)$. Experimental studies of bacterial suspensions have shown that flow velocity increases with bacterial density \cite{PhysRevLett.98.158102,wei2024ScalingTransition, perezestay2025bacteriacollectivemotionscalefree}. Motivated by these prior observations, we assume for simplicity a linear relationship, 
\begin{align}
  U_{I\to J}(t)
  &= U_0 + c\left(\frac{\rho_I(t)+\rho_J(t)}{2} - \rho_0\right) \notag\\
  &\hspace{1cm} \text{for the directed link from $I$ to $J$}, \label{eq2}
\end{align}
where $U_0$ and $c$ are model parameters and $\rho_0$ is the uniform reference density. We then apply \eqref{eq1} and \eqref{eq2} to the directional kagome network shown in \figpref{fig3}{a}. 

We first analyze the bulk band structure of this model under fully periodic boundary conditions. Linearizing \eqref{eq1} around the uniform density $\rho_0$ $(\rho_{I}=\rho_0+\delta\rho_{I})$ and applying a Fourier transform, we obtain the Schr\"{o}dinger equation in wavenumber space, $(d/dt)\delta\rho_{\bm{k}\mu}=-\mathrm{i}(H_{\bm{k}})_{\mu\nu}\delta\rho_{\bm{k}\nu}$, where $\bm{k}$ is the wavenumber, $\mu,\nu$ label the three sites in the triangular motif of the kagome network (pink regions of \figpref{fig3}{a}), and $H_{\bm{k}}$ is a $3\times 3$ non-Hermitian Hamiltonian that depends on the dimensionless parameter $c\rho_0/(2U_0)$, see Methods and Supplementary Information. Diagonalizing $H_{\bm{k}}$ yields the bulk band dispersion relations $\mathrm{Re}[\omega_{\bm{k}}]$ shown in \figpref{fig3}{b}. For $c\rho_0/(2U_0)=0$, the three bands touch and the band gap closes. In contrast, for $c\rho_0/(2U_0)>0$, gaps open between the bands throughout the Brillouin zone. In this gapped regime, each band acquires a well-defined Chern number $N_{\mathrm{Ch}}$. Using the gauge-independent method of Ref.~\cite{doi:10.1143/JPSJ.74.1674}, we find that the top, middle, and bottom bands carry Chern numbers $-1$, $0$, and $1$, respectively, see also Supplementary Fig.~4 for the corresponding Berry curvature. These results demonstrate that the bulk band structure is topologically nontrivial. 

We next investigate the emergence of the associated edge modes in a half-periodic geometry, with periodic boundary conditions along $x$ and open boundaries along $y$, see \figpref{fig3}{a} for the definition of the lattice axes. Applying a Fourier transform only along $x$, we obtain the spectra $\mathrm{Re}[\omega_{k^x}]$ shown in \figpref{fig3}{c}. For $c\rho_0/(2U_0)>0$, where the bulk bands are gapped, branches of modes appear that connect the bulk bands across the gap. To characterize the spatial structure of these modes, we compute for each eigenmode the center of mass of its density profile along $y$, see Methods for the explicit definition. As shown in \figpref{fig3}{d}, the gap-crossing modes have their mass concentrated near the bottom and top edges of the directional kagome network. This identifies the gap-crossing modes as a pair of topological edge states.
 
To examine the implications of these topological edge states for the nonlinear dynamics of our model, we perform a direct simulation of \eqref{eq1} and \eqref{eq2} in the same half-periodic geometry, starting from a small density perturbation localized at the top-right corner of the network, see \figpref{fig3}{e} and Supplementary Movie 3. In this simulation, the perturbation travels unidirectionally along the top boundary from right to left. This agrees with the sign of the group velocity of the edge modes localized at the top boundary (red modes in \figpref{fig3}{d}). We also note that the observed propagation speed is close to the group velocity and is of order $c\rho_0/2$ (here $c\rho_0/2=10$), see \figpref{fig3}{f}. We confirm, via eigenmode expansion (see Methods and Supplementary Information), that the density perturbation in this simulation is predominantly carried by the same edge modes, see \figpref{fig3}{g}. In addition, a simulation with fully open boundary conditions exhibits a density peak propagating along the edge with a comparable speed and turning around a corner, see Supplementary Fig.~5 and Movie 4. Taken together, these results demonstrate that our simple model provides a consistent topological characterization of the edge transport and edge localization in our experiments of directional kagome networks.

\vspace{5mm}

\section*{Network chirality results in edge localization}

\begin{figure*}[p!]
\centering
\includegraphics[width=\hsize]{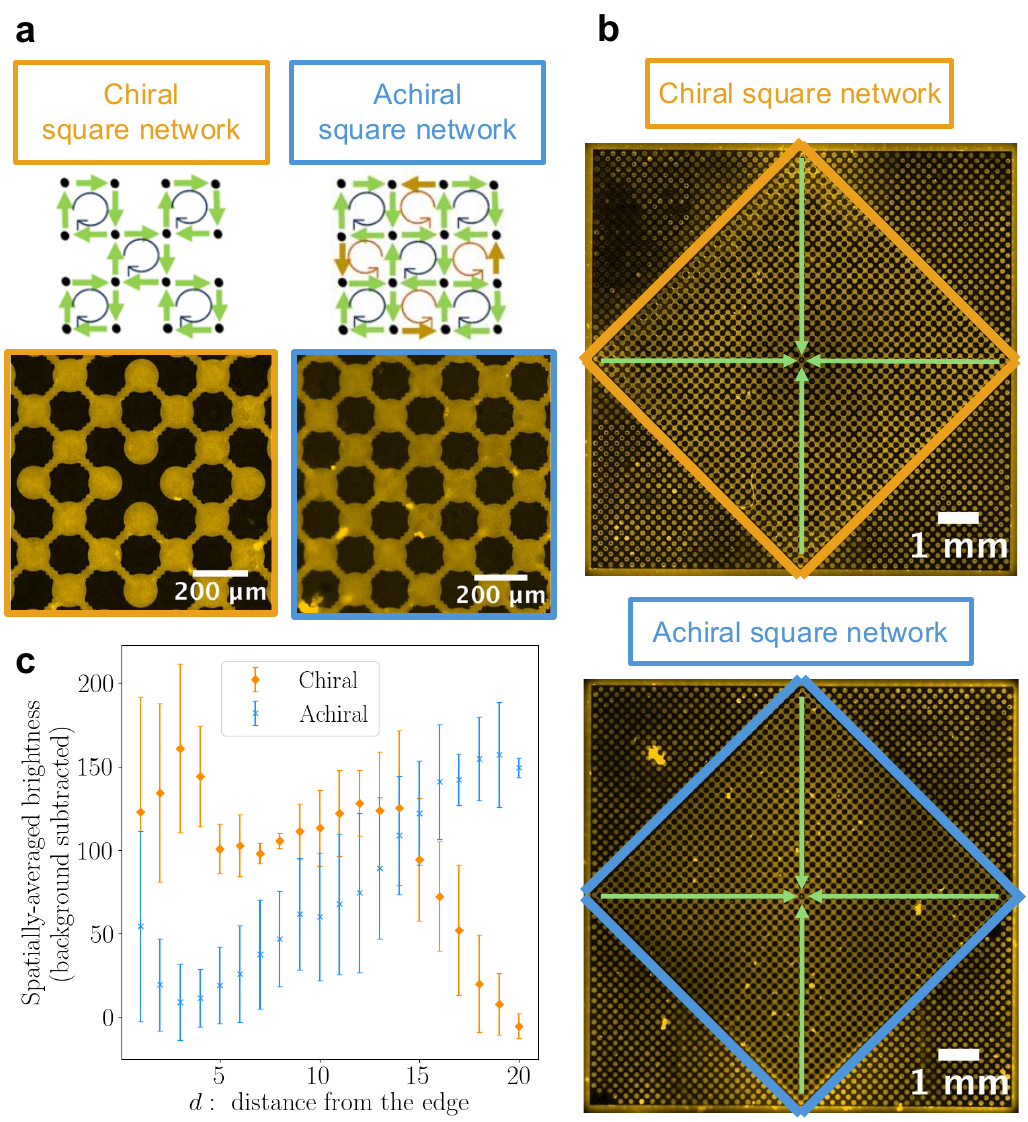}
\caption{{\bf Geometrical origin of edge states.} {\bf a}, Schematics of the chiral and achiral square networks (top), along with corresponding epifluorescence images of microfabricated structures filled with fluorescently labeled bacterial suspension (bottom). In the chiral square network, the local chirality of each square motif is preserved. In contrast, in the achiral square network, for which yellow arrows are added, negative and positive chiralities both exist and cancel each other out between neighboring square motifs. {\bf b}, Epifluorescence microscopy image of a large-scale chiral square network (top) and an achiral square network (bottom) filled with fluorescently labeled bacteria, acquired by a tiling scan. The chiral (achiral) square network is outlined in orange (blue). {\bf c}, Bacterial density profiles for chiral and achiral networks. Background-subtracted spatially-averaged brightness is plotted for the wells along the diagonal lines of the networks (green in panel {\bf b}). Error bars represent the standard error measured over four diagonal lines.}
\label{fig4}
\end{figure*}

With the emergence of topological edge states in directional kagome networks established, we now ask how such edge states can be designed in bacterial active matter through network geometry. In the quantum Hall effect, chirality plays a central role in producing unidirectional edge modes through cyclotron motion under a magnetic field. This analogy motivates us to examine whether geometrical chirality could likewise serve as a key ingredient for realizing topological edge states. 

To test this idea, we designed two types of directional square networks that differ in their global chirality, see \figpref{fig4}{a}. In the chiral square network, the local chirality of each square unit is preserved throughout the entire network. In contrast, in the achiral square network, additional directional channels (yellow arrows in \figpref{fig4}{a}) are introduced, canceling the chirality between neighboring square units. These two geometries thus provide a direct experimental means to investigate how network chirality affects the emergence of edge localization.

We then tested this geometrical effect experimentally using microfabricated chiral and achiral square networks. We measured the bacterial density distribution and plotted the background-subtracted fluorescence intensity along four diagonal lines of the networks, see \figpref{fig4}{b,c}. Edge localization was observed only in the chiral network, whereas no such localization appeared in the achiral case. These results demonstrate that network chirality governs the emergence of edge localization, establishing a geometrical route for designing and controlling edge states through geometrical chirality.

\vspace{5mm}

\section*{Conclusion and outlook}

In conclusion, we realized and controlled geometry-driven topological edge states in dense bacterial suspension. By constructing directional kagome networks composed of ratchet-shaped channels that induce strong unidirectional flows, we observed edge localization of bacterial density. Tuning the microfabricated network geometry, we further identified directional channel design and network chirality as the key geometrical ingredients governing the emergence of edge states. These findings demonstrate how concepts originally developed in quantum condensed matter physics, such as band topology and chiral edge transport, can be extended to active matter through geometrical design analogous to crystal structures.

From more general viewpoints, our results establish an environmental design strategy, where the geometry of the surrounding structure is engineered as the key ingredient that controls the emergence of topological transport, rather than relying on particle-level chirality or internal degrees of freedom. This approach provides a foundation for exploring topological transport in a broader class of active and living systems. In biological contexts, such geometry-guided flows may serve as model systems for investigating how robust transport of material and nutrients can arise in complex environments. From an applied perspective, the ability to control transport purely through network geometry offers a route toward robust flow manipulation in microfluidic, biomedical, and bioengineering devices. By integrating geometric design with the dynamics of living active matter, this approach may ultimately connect the physics of topological phases with functional transport phenomena in biological systems.

\section{Methods}

\paragraph{Preparation of dense bacterial suspension.}
Experiments were performed with a suspension of \textit{Bacillus subtilis} (strain 1085). Bacteria were first grown overnight on Luria-Bertani (LB) agar plates (BD Difco LB Broth Miller + 1.5 w\% agar) in an incubator at $\SI{37}{\degreeCelsius}$. A portion of the resulting colony was then transferred into a test tube containing $\SI{11}{mL}$ of Terrific Broth (T9179, Sigma-Aldrich). The culture was incubated at $\SI{37}{\degreeCelsius}$ and shaken at $\SI{200}{rpm}$ for 3-4 hour, until the optical density at $\SI{600}{nm}$ reached approximately 0.8. After incubation, the culture was kept shaking at room temperature ($\approx\SI{23}{\degreeCelsius}$) for at least 30 minutes to allow adaptation to the experimental condition. An aqueous solution of Rhodamine B ($\SI{1}{mg/mL}$) was then added at a dilution of 1:1000, and we waited for more than 10 minutes. Finally, the bacterial culture was concentrated about 30-fold by centrifugation at $\SI{3000}{rpm}$ ($\SI{1057}{rcf}$) for 2 minutes.

\paragraph{Microfabrication by photolithography.}
The microfabricated devices were fabricated using standard photolithography with a negative photoresist. Prior to photoresist coating, the glass substrates ($\SI{30}{mm}\times\SI{40}{mm}$, thickness No.~5, Matsunami) were treated with hexamethyldisilazane to render the surface hydrophobic and promote adhesion between the substrate and the photoresist. The negative photoresist SU-8 3050 (Kayaku Advanced) was then spin-coated onto the treated glass substrate to a thickness of approximately $\SI{150}{\micro m}$. This thickness was chosen to minimize the influence of the glass substrate on bacterial collective behavior near the liquid–air interface. Then a given network pattern was irradiated by a maskless aligner $\mu$MLA (Heidelberg Instruments). The developing process yielded the desired structures. The fabricated network comprised circular wells of radius $\SI{65}{\micro m}$ connected by channels with a mean width of $\SI{40}{\micro m}$ and length of $\SI{60}{\micro m}$. For ratchet-shaped channels, the tooth height was $\SI{8}{\micro m}$, see \figpref{fig1}{b}. The bacterial suspension was introduced into the device and sealed with a coverslip to prevent evaporation, with a $\SI{500}{\micro m}$ thick spacer placed between the glass substrate and coverslip. The sample was mounted upside-down on the microscope stage, see \figpref{fig1}{a}. Prior to pouring the suspension, we plasma-cleaned the patterned glass substrate to render the hole surfaces hydrophilic, thereby facilitating smooth injection and entry of the dense bacterial suspension.

\paragraph{Image acquisition and analysis: density distribution.} 
All experiments were conducted using an inverted microscope (IX83, Olympus) equipped with a spinning disk confocal system (Dragonfly 200, Andor), a 10× objective lens (UPLSXAPO10×, NA=0.4), and an sCMOS camera (Zyla 4.2, Andor; pixel size $\SI{6.5}{\micro m}$), with a maximum resolution of $2048\times\SI{2048}{pixels}$. To measure the steady-state bacterial density across the network, we performed epifluorescence imaging using a tiling scan. During scanning, the region of interest was set to the central $512\times \SI{512}{pixels}$ region to ensure uniform illumination at $\SI{561}{nm}$. In the case of the kagome (square) network, the final image was composed of 28 (37) tiles horizontally and 25 (37) tiles vertically, with a 2\% of overlap. The total acquisition time was approximately $\SI{5.5}{mins}$ for the kagome network and $\SI{10.5}{mins}$ for the square network. The acquired tiles were stitched using the built-in stitching function in the image acquisition software (Fusion 2.4, Andor). We then computed the average fluorescence intensity within each circular well. Here, bacteria were labeled with fluorescent dye, and the averaged brightness can be regarded as a measure of local bacterial density. To remove systematic noise, such as intensity gradients caused by slight tilting of the glass substrate, we subtracted an estimated background intensity. The background intensity was estimated from isolated wells located in the four pink dashed areas shown in Supplementary Fig.~1(a). The average brightness $I(x,y)$ of wells centered at position $(x,y)$ was fitted by a plane using the least-squares method. Supplementary Fig.~1(b) shows the spatially-averaged brightness along each diagonal line of the hexagonal-shaped kagome network, before and after background subtraction (left and right panels, respectively). After subtraction, the brightness decay appears more isotropic, indicating effective background correction. We also plotted the ratio of the spatially-averaged brightness to the estimated background intensity and confirmed that it does not affect the qualitative results.

\paragraph{Image acquisition and analysis: velocity field.} 
To capture bacterial collective flow within the microfabricated structures, we performed confocal imaging at $\SI{50}{fps}$ for 30 seconds for the triangular structures shown in Fig.~1(c) and for 60 seconds for the kagome networks shown in Fig.~1(d) and Fig.~2(a). The velocity fields were extracted using the MATLAB-based particle image velocimetry (PIV) package PIVlab \cite{Thielicke-2021}. The interrogation window was set to $16\times \SI{16}{pixels}=9.2\times \SI{9.2}{\micro m}$, with a step size of $\SI{8}{pixels}=\SI{4.6}{\micro m}$ (i.e., 50\% overlap). We then quantified the channel flow $U$ shown in \figpref{fig2}{b}, defined as a scalar constant representing the effective flow strength through each ratchet-shaped channel. As illustrated in Supplementary Fig.~3, we manually selected the region corresponding to each channel in PIVlab and computed the temporally and spatially averaged velocity vector $\bm{u}=(u_x,u_y)$ using the Plot mode. We then projected $\bm{u}$ onto the channel axis $\bm{d}$ and defined the scalar channel flow as $U=\bm{u}\cdot\bm{d}$. Here $\bm{d}$ is a unit vector along the channel centerline, pointing from the departure well to the arrival well, which defines the direction of rectified transport in the ratchet-shaped channel.

To examine whether the characteristic bacterial flow observed in the directional kagome network leads to edge localization, we analyzed the steady-state solution of the bacterial density by \eqref{eq1}, using the experimentally measured, time-averaged $U_{I\to J}$ values shown in \figpref{fig2}{b}. Because $U_{I\to J}$ constitutes a constant matrix here, we can find the steady-state density by setting \eqref{eq1} equal to zero. This steady-state solution is independent of the initial condition, as ensured by the Perron-Frobenius theorem. The solution shown in \figpref{fig2}{f} was obtained by numerically solving the corresponding eigenvalue problem represented by a $24\times 24$ matrix, whose elements are the experimentally determined $U_{I\to J}$.

\paragraph{Bacterial transport model.} 
To theoretically characterize the experimentally observed edge localization (\figpref{fig1}{e,f}, \figpref{fig2}{e,f}) and the stronger flow along the edge (\figpref{fig2}{c}) in the directional kagome network, we developed a bacterial transport model extending the framework shown in \figpref{fig2}{d}. The explicit form of \eqref{eq1} for the directional kagome network, where the position of each site $I=(i,j)\nu$ ($\nu=\alpha,\beta,\gamma$ labels the three sites within each triangular motif $(i,j)$; see \figpref{fig3}{a}) is denoted by $\bm{R}_I=(R^x_I, R^y_I)$, is given by
\begin{align}
  \frac{\partial\rho_{(i,j)\alpha}}{\partial t}=
  &-\rho_{(i,j)\alpha}\left(U_{(i,j)\alpha\to(i,j)\beta}+U_{(i,j)\alpha\to(i-1,j)\beta}\right)\notag\\  &+\rho_{(i,j)\gamma}U_{(i,j)\gamma\to(i,j)\alpha}\notag\\
  &+\rho_{(i,j-1)\gamma}U_{(i,j-1)\gamma\to(i,j)\alpha} \notag\\
  \frac{\partial\rho_{(i,j)\beta}}{\partial t}=
  &-\rho_{(i,j)\beta}\left(U_{(i,j)\beta\to(i,j)\gamma}+U_{(i,j)\beta\to(i+1,j-1)\gamma}\right)\notag\\
  &+\rho_{(i,j)\alpha}U_{(i,j)\alpha\to(i,j)\beta}\label{eq4}\\
  &+\rho_{(i+1,j)\alpha}U_{(i+1,j)\alpha\to(i,j)\beta} \notag\\
  \frac{\partial\rho_{(i,j)\gamma}}{\partial t}=&-\rho_{(i,j)\gamma}\left(U_{(i,j)\gamma\to(i,j)\alpha}+U_{(i,j)\gamma\to(i,j+1)\alpha}\right)\notag\\
  &+\rho_{(i,j)\beta}U_{(i,j)\beta\to(i,j)\gamma}\notag\\
  &+\rho_{(i-1,j+1)\beta}U_{(i-1,j+1)\beta\to(i,j)\gamma}. \notag
\end{align}

To analyze the dispersion relation for the directional kagome network, we considered the time evolution equations (3) with $U_{I\to J}$ replaced by \eqref{eq2}, and linearized them around the uniform density $\rho_0$. Then, expressing the linearized dynamics of $\delta\rho_{I}=\rho_{I}-\rho_{0,I}$ in Fourier space, we obtain the Schr\"{o}dinger equation in wavenumber space, $(d/dt)\delta\rho_{\bm{k}\mu}=-\mathrm{i}(H_{\bm{k}})_{\mu\nu}\delta\rho_{\bm{k}\nu}$. Here, $\mu$ and $\nu$ label the three sites in the triangular motif, and $H_{\bm{k}}$ is a $3\times 3$ non-Hermitian Hamiltonian that depends on the dimensionless parameter $c\rho_0/(2U_0)$. The matrix $H_{\bm{k}}$ is given by
\begin{equation}
    H_{\bm{k}}=\mathrm{i}U_0\left(\begin{array}{ccc}
        -2 & -\alpha f^{*}_2 & \beta f_1 \\
        \beta f_2 & -2 & -\alpha f^{*}_3 \\
        -\alpha f^{*}_1 & \beta f_3 & -2
    \end{array}\right), \label{eq5}
\end{equation}
with
\begin{equation}
    \alpha=\frac{c\rho_0}{2U_0}, \ \beta=1+\frac{c\rho_0}{2U_0}, \ f_{\nu}=1+e^{-2\mathrm{i}\bm{k}\cdot\bm{a}_{\nu}},
\end{equation}
where $\bm{a}_1=(0,-1), \bm{a}_2=(1,0)$ and $\bm{a}_3=(-1,1)$ are the lattice vectors. From the Schr\"{o}dinger equation, we obtain the dispersion relation by solving the eigenvalue problem, $(H_{\bm{k}})_{\mu\nu}\delta\rho_{\bm{k}\nu}=\omega_{\bm{k}}\delta\rho_{\bm{k}\mu}$. When the bands are separated by gaps throughout the Brillouin zone, the Chern number of band $n$, defined by the following equation with the eigenvector $\bm{\delta\rho}^n_{\bm{k}}$, is quantized:
\begin{equation}
    N^n_{\mathrm{Ch}}=\frac{1}{2\pi}\int_{\mathrm{BZ}}B^n_{\bm{k}}\ d\bm{k}, \label{eq6}
\end{equation}
with $B^n_{\bm{k}}=\bm{\nabla}\times\bm{A}^n_{\bm{k}}$ and $\bm{A}^n_{\bm{k}}=\mathrm{i}(\bm{\delta\rho}^n_{\bm{k}})^{\dagger}\cdot(\bm{\nabla}_{\bm{k}}\bm{\delta\rho}^n_{\bm{k}})$. The Chern numbers were calculated numerically using the gauge-independent method of Ref.~\cite{doi:10.1143/JPSJ.74.1674}, see Supplementary Information for details.

For the half-periodic geometry, with boundary conditions periodic in $x$ and open in $y$, we did Fourier transforms only in $x$, i.e., $\delta\rho_{(i,j)\nu}=\sum_{k^x}\delta\rho_{(k^x,j)\nu}\cdot\mathrm{exp}(\mathrm{i}k^xR^x_{(i,j)\nu})$. This yields the Schr\"{o}dinger equation, $(d/dt)\bm{\delta\rho}_{k^x}=-\mathrm{i}\mathcal{H}_{k^x}\bm{\delta\rho}_{k^x}$, where $\bm{\delta\rho}_{k^x}=(\delta\rho_{(k^x,j=0)\alpha},\cdots, \delta\rho_{(k^x,N_y-1)\gamma})^T$ and $\mathcal{H}_{k^x}$ is a block-tridiagonal matrix with $N_y$ blocks of the form 
\begin{equation}
\mathcal{H}_{k^x} =
\begin{pmatrix}
  A & B & 0 & \cdots & 0 \\
  B^\dagger & A & B & \ddots & \vdots \\
  0 & B^\dagger & A & \ddots & 0 \\
  \vdots & \ddots & \ddots & \ddots & B \\
  0 & \cdots & 0 & B^\dagger & A
\end{pmatrix}.
\end{equation}
The explicit forms of matrices $A$ and $B$ are provided in Supplementary Information. \figpref{fig3}{c} shows the dispersion relation computed for $N_y=21$. 

To identify edge-localized modes, we computed the $y$-coordinate of the center of mass for each eigenmode labeled by $(k^x, m)$ as follows: 
\begin{equation}
  \langle j\rangle_{k^x,m}
  =
  \frac{\displaystyle\sum_{j} j \sum_{\nu} \mathrm{Re}\left[\left(L^{(m)}_{(k^x,j)\nu}\right)^{*}R^{(m)}_{(k^x,j)\nu}\right]}
       {\displaystyle\sum_{j} \sum_{\nu} \mathrm{Re}\left[\left(L^{(m)}_{(k^x,j)\nu}\right)^{*}R^{(m)}_{(k^x,j)\nu}\right]},
\end{equation}
where $R^{(m)}_{(k^x,j)\nu}$ and $L^{(m)}_{(k^x,j)\nu}$ denote the right and left eigenvector components, respectively, on site $\nu$ within the unit triangular motif at position $y=j$ for mode $(k^x,m)$. The result is shown in \figpref{fig3}{d}. Modes with $\langle j\rangle_{k^x,m} \approx 0$ ($N_y-1$), shown in blue (red) in \figpref{fig3}{d}, are localized near the bottom (top) boundary of the directional kagome network.

\paragraph{Direct simulations of the bacterial transport model.} 

We performed direct simulations of \eqref{eq1} and \eqref{eq2} under both half-periodic boundary conditions (\figpref{fig3}{e} and Supplementary Movie 3) and fully-open boundary conditions (Supplementary Fig.~5 and Supplementary Movie 4). In all cases, we numerically solved \eqref{eq1} and \eqref{eq2} for $U_0=1$ and $c\rho_0/2=10$, and $\bm{\rho}_0=(1,\cdots,1)$, using the fourth-order Runge-Kutta method \cite{Süli_Mayers_2003} with discretized time $t_n=n\Delta t$ $(n=0,1,\cdots)$, where the time step was set to $\Delta t=0.005$. At $t=0$, the density deviation $\delta\rho_I=\rho_I-\rho_{0,I}$ was set to 0.01 at the site indicated by green in \figpref{fig3}{e} and Supplementary Fig.~5(a); at all other sites $\delta\rho_I=0$. Simulations in the half-periodic geometry were performed on a system with $N_{x}=20$ and $N_y=21$, with periodic boundary conditions along the $x$ direction implemented by identifying $\delta\rho_{(i=N_{x}, j)\nu}=\delta\rho_{(i=0, j)\nu}$, whereas simulations with fully-open boundary conditions were carried out on a system with $N_x=N_y=21$.

\paragraph{Mode expansion of the simulated density.}

To elucidate the relationship between the edge transport observed in the direct simulations of the nonlinear model and the gap-crossing modes of the half-periodic Hamiltonian $\mathcal{H}_{k^x}$, we expressed the simulated density field, $\delta\rho^{\text{sim}}_{(i,j)\nu}(t)$ $(i=0,\cdots, N_{x}-1; j=0, \cdots, N_y-1; \nu=\alpha,\beta,\gamma)$, in terms of the eigenmodes of $\mathcal{H}_{k^x}$, $\delta\rho^{(m)}_{(k^x,j)\nu}$. Specifically, we expanded $\delta\rho^{\text{sim}}_{(k^x,j)\nu}$ as follows:
\begin{equation}
	\delta\rho^{\text{sim}}_{(k^x,j)\nu}(t)=\sum_{m}c^{(m)}_{k^x}(t)\cdot\delta\rho^{(m)}_{(k^x,j)\nu}.
\end{equation}
The values of $|c^{(m)}_{k^x}(t)|$ at a representative time are shown in \figpref{fig3}{g} for all modes $(k^x,m)$. Details of the numerical procedure for evaluating $c^{(m)}_{k^x}(t)$, including the biorthogonal projection in the non-Hermitian case, are provided in Supplementary Information.

\section*{Acknowledgments}
We thank T. Yoshida (Kyoto University) for discussions on their theoretical work \cite{PhysRevE.104.025003} and for sharing the accompanying sample code for band-structure calculations. We also thank T. Sagawa, Y. Ashida (The University of Tokyo) and K. Sone (University of Tsukuba) for valuable discussions.   
This work is supported in part by JSPS KAKENHI Grant Numbers
JP19H05800, 
JP20K14426, 
JP23K25838, 
JP24K00593, 
JP25K22005,
Japan Science and Technology Agency (JST) FOREST Grant Number JPMJFR2364, 
JST PRESTO Grant Number JPMJPR21O8, 
JST SPRING Grant Number JPMJSP2108, 
and the JSPS Core-to-Core Program ``Advanced core-to-core network for the physics of self-organizing active matter (JPJSCCA20230002)''. 
Y.U. acknowledges support from FoPM, WINGS Program, the University of Tokyo.

\section*{Competing interests}
The authors declare no competing interests.

\section*{Data availability}

All the experimental and numerical data, as well as the relevant codes and scripts, are available upon request.

\section*{Author contributions}
Y.U., D.N., K.A.T. conceived the project. 
Y.U. fabricated the devices, performed the experiments, and analyzed the experimental data, with the help of D.N.
Y.U. formulated the theory and performed the numerical simulations through discussions with K.A.T.
All authors discussed and interpreted the results, and wrote the manuscript.

\bibliography{ref}

\end{document}


\onecolumngrid

\title{Supplementary Information for\\ ``Designing topological edge states in bacterial active matter''}

\author{Yoshihito Uchida}
\email{uchida@noneq.phys.s.u-tokyo.ac.jp}
\affiliation{Department of Physics,\! The University of Tokyo,\! 7-3-1 Hongo,\! Bunkyo-ku,\! Tokyo 113-0033,\! Japan}%

\author{Daiki Nishiguchi}
\email{nishiguchi@phys.sci.isct.ac.jp}
\affiliation{Department of Physics,\! Institute of Science Tokyo,\! 2-12-1 Ookayama,\! Meguro-ku,\! Tokyo 152-8551,\! Japan}%
\affiliation{Department of Physics,\! The University of Tokyo,\! 7-3-1 Hongo,\! Bunkyo-ku,\! Tokyo 113-0033,\! Japan}%

\author{Kazumasa A. Takeuchi}
\email{kat@kaztake.org}
\affiliation{Department of Physics,\! The University of Tokyo,\! 7-3-1 Hongo,\! Bunkyo-ku,\! Tokyo 113-0033,\! Japan}%
\affiliation{Universal Biology Institute,\! The University of Tokyo,\! 7-3-1 Hongo,\! Bunkyo-ku,\! Tokyo 113-0033,\! Japan}%
\affiliation{Institute for Physics of Intelligence,\! The University of Tokyo,\! 7-3-1 Hongo,\! Bunkyo-ku,\! Tokyo 113-0033,\! Japan}%

\date{\today}

\maketitle


\section{THEORETICAL ANALYSIS OF BACTERIAL TRANSPORT MODEL}

\subsection{A. Bacterial transport model in direcional kagome network}

To characterize the experimentally observed edge localization in the directional kagome network, we constructed a simplified bacterial transport model extending the framework shown in Fig.~2(d). In the model, the time evolution of bacterial density $\rho_I(t)$ at each site $I$, whose position is denoted by $\bm{R}_I=(R^x_I, R^y_I)$, is governed by 
\begin{equation}
    \frac{\partial\rho_{I}}{\partial t}=\sum_{J\in\text{in}}\rho_{J}U_{J\to I}-\sum_{J\in\text{out}}\rho_{I}U_{I\to J}, \label{SIII_eq0}
\end{equation}
where the directed channel from $I$ to $J$ is assigned directional flow strength $U_{I\to J}(t)$ that represents channel-mediated transport. The first and second terms on the right-hand side describe the inflow from neighboring sites connected by incoming channels and the outflow through outgoing channels, respectively.

We considered the directional kagome network composed of unit triangular motifs, shown in Fig.~3(a) in the main text. The lattice vectors are $\bm{a}_1=(0,-1)$, $\bm{a}_2=(1,0)$, and $\bm{a}_3=(-1,1)$, and the site positions $\bm{R}_{(i,j)\nu}$ are defined as  
\begin{equation}
  \bm{R}_{(i,j)\alpha}=2i\bm{a}_2+2j(-\bm{a}_1), \ \bm{R}_{(i,j)\beta}=\bm{R}_{(i,j)\alpha}+\bm{a}_2, \  \bm{R}_{(i,j)\gamma}=\bm{R}_{(i,j)\beta}+\bm{a}_3. \label{SIII_eq2}
\end{equation}
The time evolution equations of the bacterial density $\rho_{I}(t)$ are given by
\begin{align}
  \frac{\partial\rho_{(i,j)\alpha}}{\partial t}&=-\rho_{(i,j)\alpha}\left(U_{(i,j)\alpha\to(i,j)\beta}+U_{(i,j)\alpha\to(i-1,j)\beta}\right)+\rho_{(i,j)\gamma}U_{(i,j)\gamma\to(i,j)\alpha}+\rho_{(i,j-1)\gamma}U_{(i,j-1)\gamma\to(i,j)\alpha}, \notag\\
  \frac{\partial\rho_{(i,j)\beta}}{\partial t}&=-\rho_{(i,j)\beta}\left(U_{(i,j)\beta\to(i,j)\gamma}+U_{(i,j)\beta\to(i+1,j-1)\gamma}\right)+\rho_{(i,j)\alpha}U_{(i,j)\alpha\to(i,j)\beta}+\rho_{(i+1,j)\alpha}U_{(i+1,j)\alpha\to(i,j)\beta}, \label{SIII_eq1}\\
  \frac{\partial\rho_{(i,j)\gamma}}{\partial t}&=-\rho_{(i,j)\gamma}\left(U_{(i,j)\gamma\to(i,j)\alpha}+U_{(i,j)\gamma\to(i,j+1)\alpha}\right)+\rho_{(i,j)\beta}U_{(i,j)\beta\to(i,j)\gamma}+\rho_{(i-1,j+1)\beta}U_{(i-1,j+1)\beta\to(i,j)\gamma}. \notag
\end{align}

To solve \eqsref{SIII_eq1}, we specify the relationship between $\rho_{(i,j)\nu}$ and $U_{(i,j)\nu\to (i',j')\nu'}$. Motivated by previous experiments on bacterial suspensions showing that the flow velocity increases with bacterial density \cite{PhysRevLett.98.158102,wei2024ScalingTransition, perezestay2025bacteriacollectivemotionscalefree}, here we assume a linear relationship between the channel flow and local bacterial density. For the channel directed from site $(i,j)\nu$ to site $(i',j')\nu'$, 
\begin{align}
  U_{(i,j)\nu\to (i',j')\nu'}&=U_0+c\left(\frac{\rho_{(i,j)\nu}+\rho_{(i',j')\nu'}}{2}-\rho_0\right), \label{SIII_eq3}
\end{align}
where $U_0$ and $c$ are parameters and $\rho_0$ denotes the uniform reference density.

We first analyzed the bulk band topology. Linearizing \eqsref{SIII_eq1} around the uniform density $\rho_{(i,j)\nu}=\rho_0+\delta\rho_{(i,j)\nu}$, we obtain 
\begin{align}
  \frac{\partial}{\partial t}\delta\rho_{(i,j)\alpha}&=U_0(-2\delta\rho_{(i,j)\alpha}+\delta\rho_{(i,j-1)\gamma}+\delta\rho_{(i,j)\gamma})+\frac{c\rho_0}{2}(-\delta\rho_{(i,j)\beta}-\delta\rho_{(i-1,j)\beta}+\delta\rho_{(i,j-1)\gamma}+\delta\rho_{(i,j)\gamma}), \notag\\
  \frac{\partial}{\partial t}\delta\rho_{(i,j)\beta}&=U_0(-2\delta\rho_{(i,j)\beta}+\delta\rho_{(i+1,j)\alpha}+\delta\rho_{(i,j)\alpha})+\frac{c\rho_0}{2}(-\delta\rho_{(i,j)\gamma}-\delta\rho_{(i+1,j-1)\gamma}+\delta\rho_{(i+1,j)\alpha}+\delta\rho_{(i,j)\alpha}), \label{SIII_eq4}\\
  \frac{\partial}{\partial t}\delta\rho_{(i,j)\gamma}&=U_0(-2\delta\rho_{(i,j)\gamma}+\delta\rho_{(i-1,j+1)\beta}+\delta\rho_{(i,j)\beta})+\frac{c\rho_0}{2}(-\delta\rho_{(i,j)\alpha}-\delta\rho_{(i,j+1)\alpha}+\delta\rho_{(i-1,j+1)\beta}+\delta\rho_{(i,j)\beta}). \notag
\end{align}
Notably, when $U_0=0$, \eqsref{SIII_eq4} reduce to the evolutionary dynamics model on a kagome network considered in Ref.~\cite{PhysRevE.104.025003}, which also shows topological edge states characterized by nontrivial Chern numbers. From \eqsref{SIII_eq4}, we apply the Fourier transform, 
\begin{equation}
  \delta\rho_{(i,j)\nu}=\sum_{\bm{k}}\delta\rho_{\bm{k}\nu}\cdot\mathrm{exp}(\mathrm{i}\bm{k}\cdot\bm{R}_{(i,j)\nu}), \label{SIII_eq40}
\end{equation}
and obtain the Schr\"{o}dinger equation $(d/dt)\bm{\delta\rho}_{\bm{k}}=-\mathrm{i}H_{\bm{k}}\bm{\delta\rho}_{\bm{k}}$, where $\bm{\delta\rho}_{\bm{k}}=(\delta\rho_{\bm{k}\alpha}, \delta\rho_{\bm{k}\beta}, \delta\rho_{\bm{k}\gamma})$, and $H_{\bm{k}}$ is the $3\times 3$ matrix corresponding to Hamiltonian. The matrix $H_{\bm{k}}$ is given by 
\begin{equation}
    H_{\bm{k}}=\mathrm{i}U_0\left(\begin{array}{ccc}
        -2 & -\alpha f^{*}_2 & \beta f_1 \\
        \beta f_2 & -2 & -\alpha f^{*}_3 \\
        -\alpha f^{*}_1 & \beta f_3 & -2
    \end{array}\right), \label{SIII_eq41}
\end{equation}
with 
\begin{equation}
    \alpha=\frac{c\rho_0}{2U_0}, \ \beta=1+\frac{c\rho_0}{2U_0}, \ f_{\nu}=1+e^{-2\mathrm{i}\bm{k}\cdot\bm{a}_{\nu}}. \label{SIII_eq42}
\end{equation}
The real and imaginary parts of the eigenvalues of $H_{\bm{k}}$ correspond to the oscillation frequency and growth rate of the collective density dynamics. The dispersion relations of this model are shown in Fig.~3(b) in the main text. Notably, when $c>0$, a band gap opens. We then evaluated the Berry curvature, 
\begin{equation}
    B^n_{\bm{k}}=\bm{\nabla}\times\bm{A}^n_{\bm{k}}, \ \bm{A}^n_{\bm{k}}=\mathrm{i}(\bm{\delta\rho}^n_{\bm{k}})^{\dagger}\cdot(\bm{\nabla}_{\bm{k}}\bm{\delta\rho}^n_{\bm{k}}), \label{SIII_eq44}
\end{equation}
and the resulting Chern number
\begin{equation}
    N^n_{\mathrm{Ch}}=\frac{1}{2\pi}\int_{\mathrm{BZ}}B^n_{\bm{k}}\ d\bm{k}, \label{SIII_eq43}
\end{equation}
using the gauge-independent method of Ref.~\cite{doi:10.1143/JPSJ.74.1674}. We find that the top, middle, and bottom bands carry Chern numbers
of $-1$, $0$, and $+1$, respectively. These values indicate that chiral edge modes should emerge within the bulk band gaps, when an open boundary is introduced.

To confirm this, we performed a band structure calculation under a half-periodic geometry, where periodicity is imposed only along the $x$-direction. The corresponding Fourier transform is:
\begin{equation}
  \delta\rho_{(i,j)\nu}=\sum_{k^x}\delta\rho_{(k^x,j)\nu}\cdot\mathrm{exp}(\mathrm{i}k^xR^x_{(i,j)\nu}). \label{SIII_eq5}
\end{equation}
Then we obtained the Schr\"{o}dinger equation, $(d/dt)\bm{\delta\rho}_{k^x}=-\mathrm{i}\mathcal{H}_{k^x}\bm{\delta\rho}_{k^x}$, with
\begin{equation}
    \bm{\delta\rho}_{k^x}=(\delta\rho_{(k^x,j=0)\alpha}, \delta\rho_{(k^x,j=0)\beta}, \delta\rho_{(k^x,j=0)\gamma}, \cdots, \delta\rho_{(k^x,j=N_y-1)\alpha}, \delta\rho_{(k^x,j=N_y-1)\beta}, \delta\rho_{(k^x,j=N_y-1)\gamma})^T, 
\end{equation}
where $\mathcal{H}_{k^x}$ is a block tridiagonal matrix composed of $N_y$ blocks 
\begin{equation}
\mathcal{H}_{k^x} = U_0
\begin{pmatrix}
  A & B & 0 & \cdots & 0 \\
  B^\dagger & A & B & \ddots & \vdots \\
  0 & B^\dagger & A & \ddots & 0 \\
  \vdots & \ddots & \ddots & \ddots & B \\
  0 & \cdots & 0 & B^\dagger & A
\end{pmatrix}, \label{SIII_eq10}
\end{equation}
with the matrices
\begin{equation}
  A=\mathrm{i}\left(
    \begin{array}{ccc}
      -2 & -\alpha(1+e^{2\mathrm{i}k^x}) & \beta\\
      \beta(1+e^{-2\mathrm{i}k^x}) & -2 & -\alpha\\
      -\alpha & \beta & -2
    \end{array}
  \right), \label{SIII_eq11}
\end{equation}
\begin{equation}
  B=\mathrm{i}\left(
    \begin{array}{ccc}
      0 & 0 & 0\\
      0 & 0 & 0\\
      -\alpha & \beta e^{2\mathrm{i}k^x} & 0
    \end{array}
  \right). \label{SIII_eq12}
\end{equation}
Figure 3(c) in the main text shows the dispersion relation for $N_y=21$.

\subsection{B. Mode expansion of the simulated density}

Here we describe how we compute the expansion coefficients $c^{(m)}_{k^x}(t)$ shown in Fig.~3(g) in the main text. We used the density deviation field $\delta\rho^{\text{sim}}_{(i,j)\nu}(t)$ obtained by the direct simulation of the nonlinear model in the half-periodic geometry, see Methods. More specifically, we used its Fourier transform  
\begin{equation}
	\delta\rho^{\text{sim}}_{(k^x,j)\nu}(t)=\frac{1}{\sqrt{N_{x}}}\sum_{i=0}^{N_{x}-1}\delta\rho^{\text{sim}}_{(i,j)\nu}(t)\cdot \mathrm{exp}(-\mathrm{i}k^x i).
\end{equation}
For each $k^x$, the half-periodic Hamiltonian $\mathcal{H}_{k^x}$ is diagonalized to obtain its eigenvalues $\omega^{(m)}_{k^x}$ and right eigenvectors $\delta\rho^{(m)}_{(k^x,j)\nu}$ $(m=1,\cdots,3N_y)$. Collecting these eigenvectors as the columns of a matrix $V_{k^x}$, the simulated density at fixed $k^x$ and time $t$ can be written in vector form as $\bm{\delta\rho}^{\text{sim}}_{k^x}(t)=V_{k^x}\bm{c}_{k^x}(t)$, where $\bm{\delta\rho}^{\text{sim}}_{k^x}(t)$ stacks $\delta\rho^{\text{sim}}_{(k^x,j)\nu}(t)$ over $(j,\nu)$, and $\bm{c}_{k^x}(t)$ stacks the expansion coefficients $c^{(m)}_{k^x}(t)$ over $m$. Assuming that the eigenvectors form a complete basis, the coefficients are obtained by inverting this linear relation, $\bm{c}_{k^x}(t)=V^{-1}_{k^x}\bm{\delta\rho}^{\text{sim}}_{k^x}(t)$. In component form, this corresponds to the expansion
\begin{equation}
    \delta\rho^{\text{sim}}_{(k^x,j)\nu}(t)=\sum_{m}c^{(m)}_{k^x}(t)\cdot\delta\rho^{(m)}_{(k^x,j)\nu}.
\end{equation}
In Fig.~3(g) of the main text, we plot the magnitude $|c^{(m)}_{k^x}(t)|$ at $t=0$, which quantifies the contribution of each eigenmode in the simulated edge transport in the nonlinear model.

\section{Supplementary Movie Captions}

\begin{description}
\item[Movie 1]
Confocal observation of dyed bacterial suspension confined within six triangular structures. The left and right pairs are connected by ratchet-shaped channels, whereas the central pair is connected by straight channels. The right pair corresponds to Fig.~1(c) in the main text. The movie is played at real-time speed. The scale bar represents $\SI{100}{\micro m}$.
\item[Movie 2]
Confocal observation of dyed bacterial suspension confined within the directional kagome network composed of 24 circular-shaped wells, corresponding to Fig.~1(d) and Fig.~2(a) in the main text. The movie is played at real-time speed. The scale bar represents $\SI{100}{\micro m}$.
\item[Movie 3]
Time evolution of the simulated bacterial density obtained by direct numerical integration of Eqs.~(1) and (2) in the main text under half-periodic geometry ($N_x=20, N_y=21$), where periodic boundary conditions are applied only in the $x$ direction, with $U_0=1$ and $c\rho_0/2=10$. At $t=0$, the density deviation $\delta\rho_I=\rho_I-\rho_{0,I}$ is set to 0.01 only at the site located at the top-right corner, with $\bm{\rho}_0=(1,\cdots,1)$; at all other sites $\delta\rho_I=0$. The simulation was performed with a time step $\Delta t=0.005$, and frames were recorded every 20 steps to generate the video. The color encodes the local density deviation $\delta\rho_I(t)$.
\item[Movie 4]
Time evolution of the simulated bacterial density obtained by direct numerical integration of Eqs.~(1) and (2) in the main text under fully-open boundary conditions ($N_x=N_y=21$) with $U_0=1$ and $c\rho_0/2=10$. At $t=0$, the density deviation $\delta\rho_I=\rho_I-\rho_{0,I}$ is set to 0.01 only at the site located at the bottom-left corner, with $\bm{\rho}_0=(1,\cdots,1)$; at all other sites $\delta\rho_I=0$. The simulation was performed with a time step $\Delta t=0.005$, and frames were recorded every 20 steps to generate the video. The color encodes the local density deviation $\delta\rho_I(t)$.
\end{description}

\bibliography{ref}

\newpage

\section{Supplementary Figures}

\begin{figure}[b!]
\includegraphics[width=\hsize,clip]{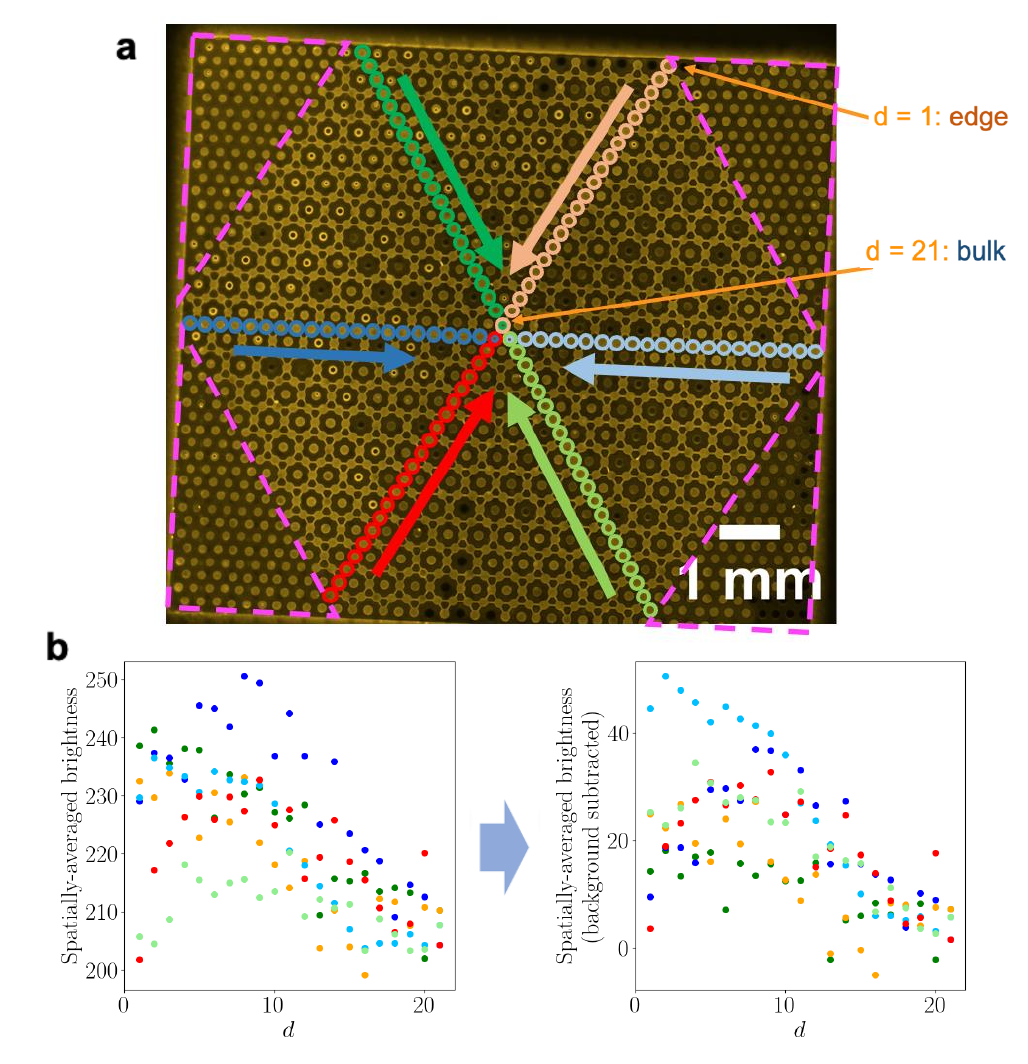}
\centering
\caption{
{\bf Measurement of bacterial density profiles in a directional kagome network.} {\bf (a)} Epifluorescence image of the entire directional kagome network. Average fluorescence intensity was measured within circular wells along six diagonal lines, highlighted with colored circles. Background intensity was estimated from isolated wells located in the pink-highlighted regions, see Methods. {\bf (b)} Average fluorescence intensity before (left panel) and after (right) background subtraction. The color of each plot corresponds to the color of the highlighted wells in panel (a). 
}
\label{figS1}
\end{figure}

\begin{figure}[p!]
\includegraphics[width=\hsize,clip]{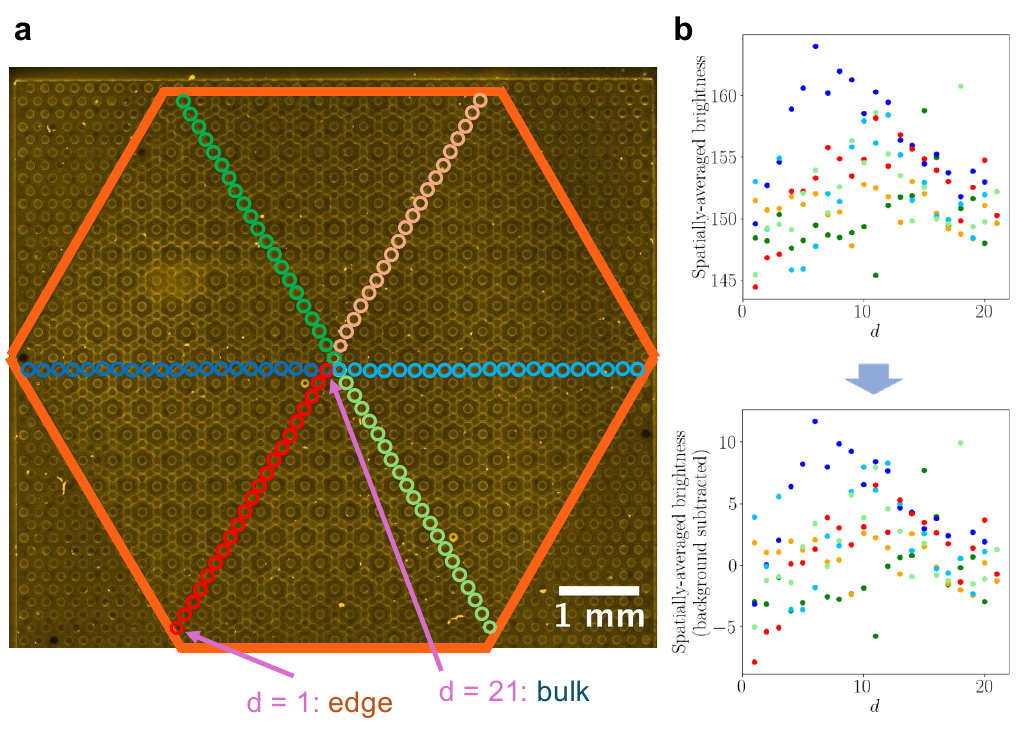}
\centering
\caption{
{\bf Measurement of bacterial density profiles in a non-directional kagome network.} {\bf (a)} Epifluorescence image of the entire non-directional kagome network. Average fluorescence intensity was measured within circular wells along six diagonal lines, highlighted with colored circles. {\bf (b)} Average fluorescence intensity before (top panel) and after (bottom) background subtraction. The color of each plot corresponds to the color of the highlighted wells in panel (a). 
}
\label{figS2}
\end{figure}

\begin{figure}[p!]
\includegraphics[width=\hsize,clip]{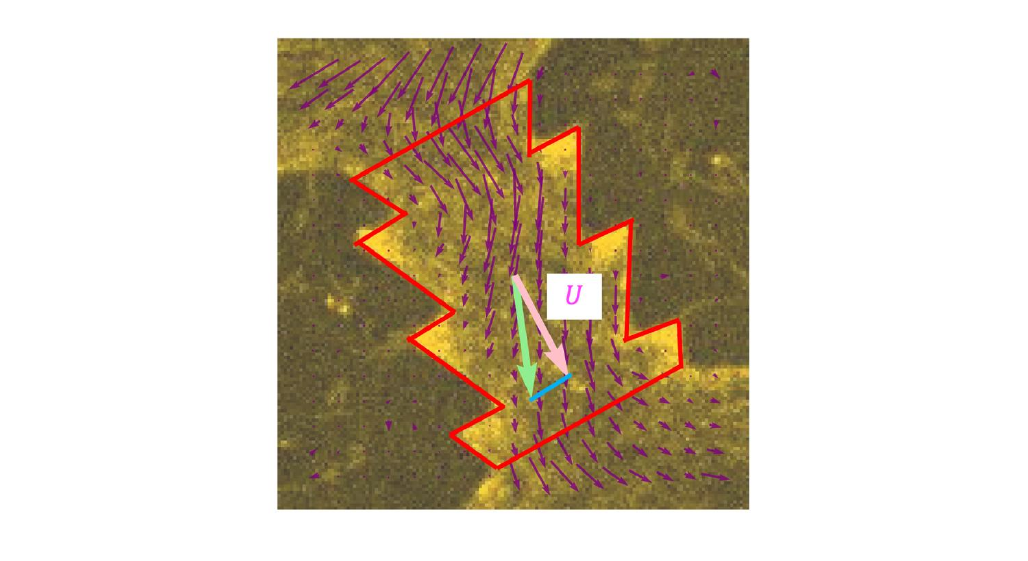}
\centering
\caption{
{\bf Evaluation of the channel flow.} Purple arrows represent the time-averaged velocity field. The green arrow indicates the temporally and spatially averaged velocity within the channel, which was obtained by manually selecting the channel region (red outline) and using the Plot mode in PIVlab \cite{Thielicke-2021}. The green vector was then projected onto the channel axis. For clarity, the lengths of the green and pink arrows are scaled to be five times the lengths of the purple arrows.
}
\label{figS3}
\end{figure}

\begin{figure}[b!]
\includegraphics[width=\hsize,clip]{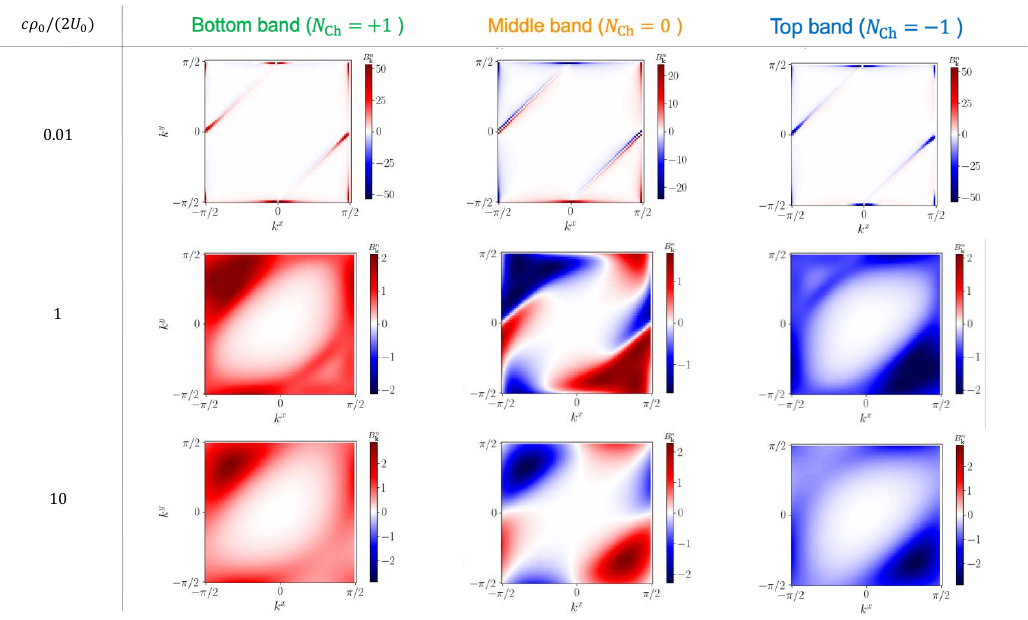}
\centering
\caption{{\bf Berry curvature of the bacterial transport model on directional kagome networks for the fully periodic case.}}
\label{figS4}
\end{figure}

\begin{figure*}[b!]
\centering
\includegraphics[width=\hsize]{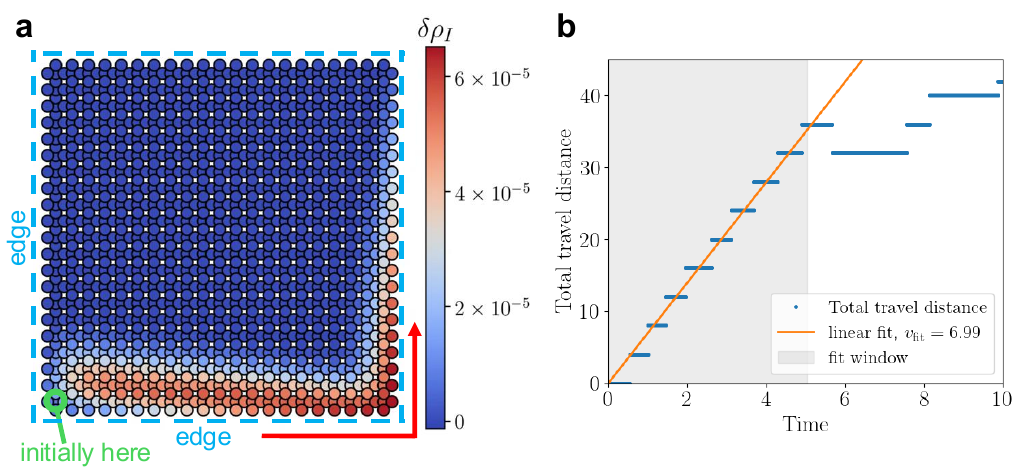}
\caption{{\bf Direct simulation of the nonlinear model under fully-open boundary conditions for $U_0=1$ and $c\rho_0/2=10$.} {\bf (a)} Snapshot at time $t=10$ from the simulation. The color indicates the density deviation $\delta\rho_I=\rho_I-\rho_{0,I}$. At $t=0$, the density deviation $\delta\rho_I$ is set to 0.01 at the left-bottom corner, marked in green, and $\delta\rho_I=0$ at all other sites. The red arrow indicates the direction of the observed edge transport. See also Supplementary Movie 4. {\bf (b)} Total travel distance $D(t)$ of the density peak. The travel distance is defined as the distance from the triangular motif where the initial density deviation was applied (marked in green in {\bf (a)}). The orange line shows a linear fit, $D(t)=v_{\text{fit}} t$. The fit is performed over the time range $t<5$ (shaded in gray), where the position of density peak moves approximately at a constant speed. From this fit, we obtain $v_{\text{fit}}=7.0\pm 1.4$.}
\label{figS5}
\end{figure*}

